 \documentclass[final,5p,times,onecolumn,authoryear]{elsarticle}

\usepackage{amssymb}

\onecolumn
\journal{New Astronomy}

\begin{document}

\begin{frontmatter}


\onecolumn
\title{Spectro-timing analysis of Be X-ray pulsar SMC X-2 during the 2022 outburst}

\onecolumn
\author[Tobrej et al.]{
Mohammed Tobrej,$^{1}$\thanks{tabrez.md565@gmail.com}
Binay Rai,$^{1}$\thanks{binayrai21@gmail.com}
Manoj Ghising,$^{1}$\thanks{manojghising26@gmail.com}
Ruchi Tamang,$^{1}$\thanks{ruchitamang76@gmail.com}
Bikash Chandra Paul$^{1}$\thanks{bcpaul@associates.iucaa.in}
\\
$^{1}$Department of Physics, North Bengal University, Siliguri, Darjeeling, WB, 734013, India
\\
}

\begin{abstract}
We present broadband X-ray observations of the High Mass X-ray Binary (HMXB) pulsar SMC X-2, using concurrent NuSTAR and NICER observations during its 2022 outburst. The source is found to be spinning with a period of 2.37281(3) s. We confirm the existence of cyclotron resonant scattering feature (CRSF) at $\sim$ 31 keV in addition to the iron emission line in the X-ray continuum of the source. Spectral analysis performed with the physical bulk and thermal Comptonization model indicates that the bulk Comptonization dominates the thermal Comptonization. Using phase-resolved spectroscopy, we have investigated the variations of the spectral parameters relative to pulse phase that may be due to the complex structure of magnetic field of the pulsar or the impact of the emission geometry. It is observed that the spectral parameters exhibit significant variabilities relative to the pulsed phase. Time-resolved spectroscopy is employed to examine the evolution of the continuum and changes in the spectral characteristics. Measurements of luminosity along with variations in cyclotron line energy and photon index suggest that the source may be accreting in the super-critical regime.
\end{abstract}



\begin{keyword}
 accretion discs-stars \sep neutron-pulsars \sep X-ray binaries \sep SMC X-2  



\end{keyword}

\end{frontmatter}



*Corresponding author

 E-mail: bcpaul@associates.iucaa.in (Bikash Chandra Paul)
\onecolumn 

\section{Introduction}
The X-ray pulsar  SMC X-2 is a bright transient in the Small Magellanic Cloud (SMC). The SMC is a Milky Way satellite located at a distance of 62 kpc \citep{Haschke}. This galaxy is rich in Be X-ray binaries, harboring a neutron star (NS) orbiting around an O-type Be companion \citep{Coee}. The non-supergiant OB spectral type star that is the optical companion in BeXRBs exhibits both infrared excess and Balmer series emission lines \citep{Reig}. The Be star rotates rapidly at velocities more than three-fourths of the Keplerian limit, resulting in an equatorial circumstellar disc \citep{Porter}
2003). The compact object in the system accretes directly from the Be-circumstellar disc. BeXRBs exhibit two types of X-ray outbursts. Type-I outbursts are known to last for a few days to several weeks \citep{Stella}, and are characterized by luminosities $\leq 10^{37}$ ergs s$^{-1}$. These are known to occur close to the periastron passage of the binary system. Type-II outbursts are giant outbursts that typically last for several weeks to months. SMC X-2 was discovered by the SAS- X-ray observatory in 1977. At the time of its discovery, a luminosity of 8.4 $\times 10 ^{37}$ ergs s$^{-1}$ in 2-11 keV energy band, considering a distance of 65 kpc \citep{clarkg,clark} was reported. Since its discovery, it has been observed by various observatories like HEAO, Einstein and ROSAT, which has inferred about the transient nature of the source \citep{Marshall, Seward, Kahabka}. The source is known to be pulsating with a period of $\sim 2.37$ s which was discovered using the RXTE and ASCA observations in 2000 \citep{Torii, Corbet, Yokogawa}. The optical companion of the source was discovered by \cite{Crampton} and was resolved into two early spectral type stars \citep{Schmidtke}. The two stars were monitored using the Optical Gravitational Lensing Experiment (OGLE) \citep{Udalski} showing the southern, fainter star to be almost constant, and the northern star to be associated  with a periodic variability $\sim$ 18.62 $\pm$ 0.02 d \citep{Schurch}. The periodic modulation estimated using RXTE measurements revealed the period at P = 18.38 $\pm$ 0.02 d \citep{Townsend} suggesting suggesting the northern  O9.5 III-V emission line star \citep{McBride}, as the actual counterpart and the observed periodicity denoting the orbital period of the binary system. \cite {jais} studied the source recently and found a clear signature of the orbital motion which is consistent with orbital period reported  in the literature using RXTE data \citep{Townsend} and optical data \citep{Roy} has been reported. An almost circular orbit with an eccentricity (e) of approximately 0.07 was determined using the RXTE data. The 2022 monitoring data indicates the orbit to be somewhat eccentric [e=0.27] \citep{jais}.
 
The system encountered outbursts in October 1977 \citep{Marshall, Seward}, January–April 2002 \citep{Corbet} 2001), August–November 2015 \citep{Kenneaa, Coee, Jaisawal}, and June–August 2022 \citep{Coe}. In 2015, the luminosity reached up to a maximum of $\sim 10^{38}$ ergs s$^{-1}$ \citep{Jaisawal, La Palombara, Lutovinov, Roy}. Since 2015, no reports of major X-ray activity had been detected in the source. Recently, an X-ray outburst was detected in June 2022 by S-CUBED \citep{Kennea} which was supported by other follow-up observations \citep{Coe}. We have studied the spectral and temporal features of the source using the Nuclear Spectroscopy Telescope Array (NuSTAR) and Neutron Star Interior Composition Explorer (NICER) observations during the 2022 outburst. 

We present a detailed study of spectral and temporal properties of SMC X-2 using simultaneous NuSTAR and NICER observations. The evolution of the spectral parameters relative to phase, time and flux/luminosity corresponding to the 2022 outburst have been discussed.

\section{Observation and Data reduction}
The data reduction for SMC X-2 has been carried out using HEASOFT v6.30. The source SMC X-2 was observed simultaneously by NuSTAR and NICER observatories on 13-07-2022.

\subsection{NuSTAR}
NuSTAR is a NASA space-based X-ray telescope, also known as SMEX-11 and Explorer 93, is used to examine high-
energy X-rays from astrophysical sources. NuSTAR is known to be the first hard X-ray focusing telescope and is sensitive in the (3-79) keV energy range. It is comprised of two identical X-ray telescope modules equipped with independent mirror systems. Each telescope is assigned with its own focal plane module FPMA and FPMB consisting of a pixelated solid state CdZnTe detector \citep{15}. The detectors are characterized by a spectral resolution of 400 eV (FWHM) at energy 10 keV. The presence of a unique multilayered coating of the grazing-incidence Wolter-I optics helps in providing X-ray imaging with an angular resolution of 18" (FWHM) \& 58" (HPD). Clean event files were created using the mission-specific NUPIPELINE. The image was observed using astronomical imaging and data visualization application DS9. We utilized the NuSTAR calibration database (CALDB) version 20220525 for performing the analysis. Spectra and light curves have been extracted by considering a circular region of $80^{\prime \prime}$ as the source region while backgrounds have been generated from an $80^{\prime \prime}$ region taken away from the source. The source and background files were utilized for obtaining the required light curve and the spectra by implementing the mission-specific task NUPRODUCTS. The background correction for the light curve was carried out using FTOOL LCMATH. The NuSTAR data has been barycentered to the solar system frame using the FTOOL BARYCORR. The spectra obtained were fitted in XSPEC version 12.12.1 \citep{s}. Based upon the ephemeris of \cite{K. L. Li}, the NuSTAR observation under consideration covers the orbital phases (0.88-0.93).

\subsection{NICER}
We have also utilized the NICER data for spectral analysis in the soft energy band. NICER is an International Space Station (ISS) external payload \citep{Gendreau} designed for studying neutron stars by soft X-ray timing. We have utilized X-ray Timing Instrument (XTI), which operates in the  (0.2-12.0) keV energy range. The standard data screening and reduction has been carried out using NICERDAS v9 and CALDBv xti20210720. The background has been estimated using the tool nibackgen3C50 v6 \citep{Remillard}. The clean event files were filtered using  NICERL2. The filtered event files were loaded to XSELECT for extracting the required light curves and pha files. Response and arf files were generated using mission-specific commands NICERARF and NICERRMF. Using the cleaned event files, the light curve was generated with a binning of 0.01s and was subsequently barycentered to the solar system frame using FTOOL BARYCORR.

\begin{table*}
\begin{center}
\begin{tabular}{ccccc}
\hline
Observatory & Observation ID &Start and stop time (MJD) &  Exposure (s)  &\\	
\hline
NuSTAR	&	90801319002	& 59773 (20:17:41:485) - 59774 (16:01:20:082)     	& 42978 \\
NICER  &     5202830119 & 59773 (07:58:54:732) - 59773 (08:37:50:732)      & 1195 \\
\hline
\end{tabular}

\caption{NuSTAR and NICER observations indicated by the observation ID, time of observation and exposure.}  
\end{center}
\end{table*} 

\section{Timing Analysis}
\begin{figure}
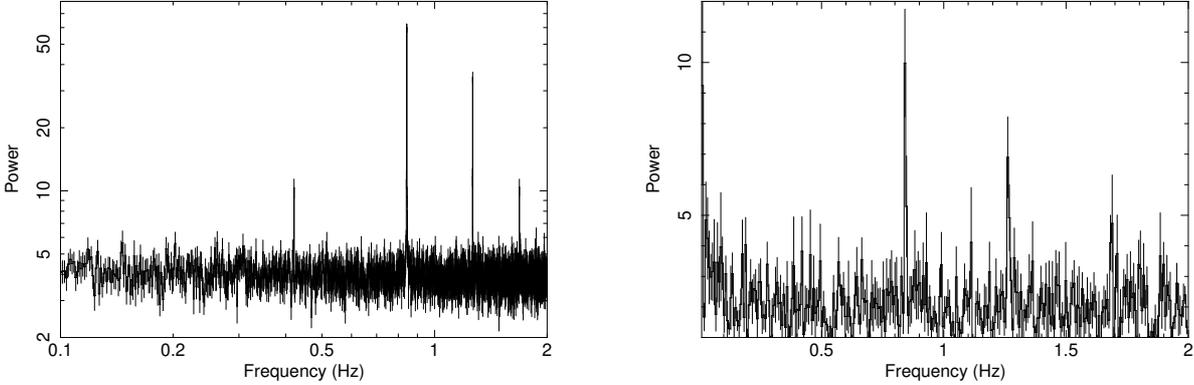

\begin{minipage}{0.35\textwidth}
\includegraphics[angle=270,scale=0.3]{powspec}
\end{minipage}
\hspace{0.10\linewidth}
\begin{minipage}{0.35\textwidth}
\includegraphics[angle=270,scale=0.3]{pspec}
\end{minipage}
\caption{\textit{Left}: FFT of the NuSTAR light curve  of the source indicating the presence of harmonics  along with the fundamental signal. Prominent peaks correspond to frequencies 0.84 Hz, 1.69 Hz, 3.33 Hz \& 6.66 Hz respectively. \textit{Right}: FFT of the NICER light curve of the source. The x-axis has been plotted in log scale.}
\end{figure}

\begin{figure}
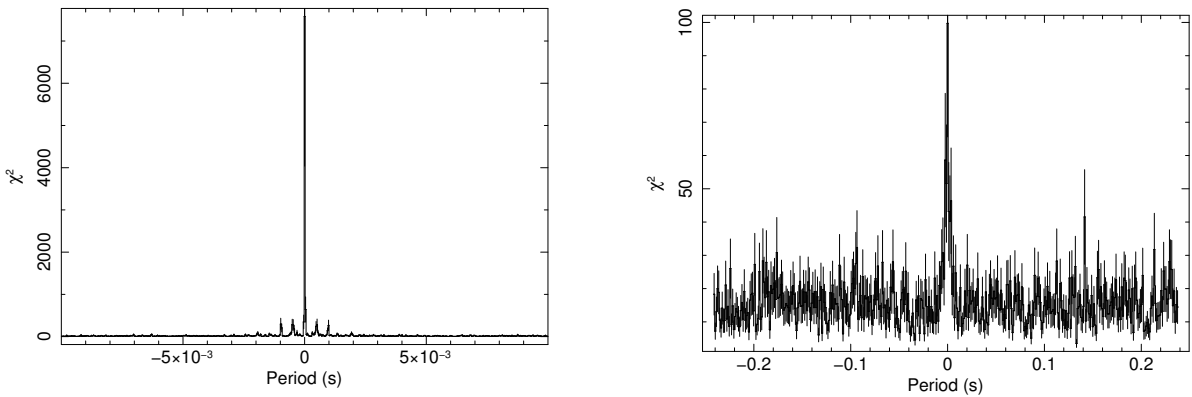

\begin{minipage}{0.35\textwidth}
\includegraphics[angle=270,scale=0.3]{fft_nustar}
\end{minipage}
\hspace{0.10\linewidth}
\begin{minipage}{0.35\textwidth}
\includegraphics[angle=270,scale=0.3]{fft_nicer}
\end{minipage}
\caption{Periodogram of the source for NuSTAR observation \textit{(left)} and NICER observation \textit{(right)}. A coherent signal is detected at $\sim$ 2.37281 s and $\sim$ 2.37279 s for NuSTAR and NICER observations respectively.}
\end{figure}

To examine the temporal characteristics of the source, we took into account NuSTAR light curves with a 0.01 s binning. With the help of FTOOL LCURVE, light curves have been produced. Utilizing the Fast Fourier Transform (FFT) of the light curve, the pulse period was approximated.  In order to determine the pulse period, we used the epoch-folding method described in (Davies 1990; Larsson 1996), which maximises $\chi^{2}$. The tool EFSEARCH  is used to search periodicities in a time series by folding the data over a range of periods around the approximate period and by  searching  for  a  maximum $\chi^{2}$ as a function of period. As a result, we estimated that 2.37281 (3) s is the optimal period of the source. For calculating the uncertainty in the measurement of pulsations, we applied the technique outlined in  \cite{18}. The errors were estimated for each light curve using the  bootstrap method. We generated 1000 simulated light curves by adjusting the count rate in each time bin using the formula: $r_{k'} = r_{k} + \gamma\sigma_{{r_k}}$. Here,  $r_{k'}$ represents the count rates in the new light curve at the kth point and $r_{k}$ is the count rate in the original light curve, $\gamma$ is a uniformly distributed quantity between -1 and 1, and $\sigma_{{r_k}}$ is the measurement error of flux for the kth point. We applied the epoch folding technique to determine the best period for each of these 1000 samples. By averaging the best periods and calculating their standard deviation, we obtained the spin period to be 2.37281 (3) s seconds. The EFSEARCH tool provides the optimal pulse period but does not provide the folded pulse profile of the source. Consequently, the source's folded pulse profile is obtained using the tool EFOLD. It is evident from Figure 1 that the FFT of the NuSTAR data is associated with substantial harmonic peaks at $\sim$ 1.19, 0.59, 0.30, and 0.15 seconds respectively. The NICER light-curve was analyzed in a similar manner and the corresponding pulse period was estimated to be 2.37279 (4) s. The periodograms corresponding to the NuSTAR and NICER observation are presented in Figure 2.
\begin{figure}
\begin{minipage}{0.35\textwidth}
\includegraphics[angle=0,scale=0.3]{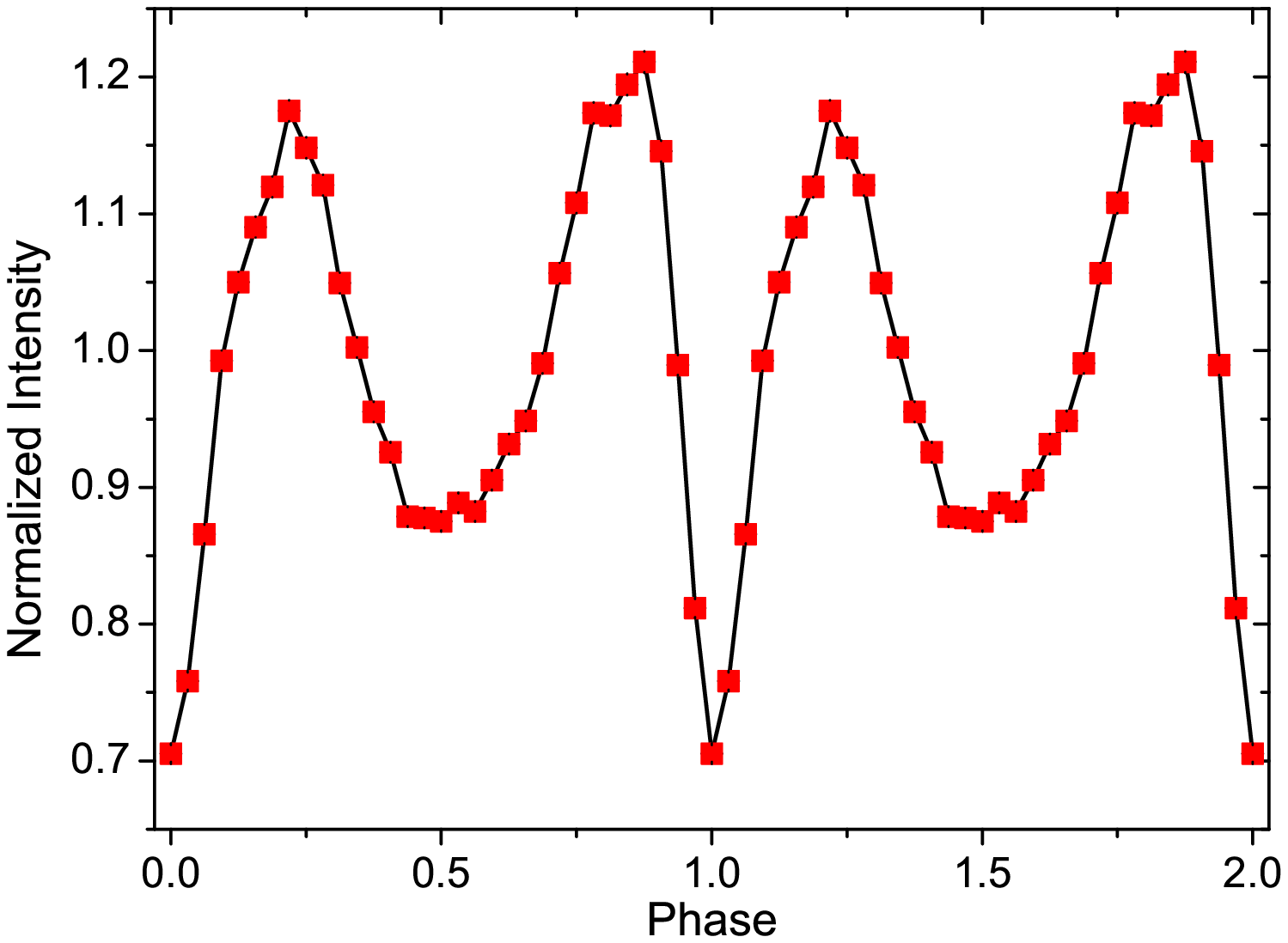}
\end{minipage}
\hspace{0.10\linewidth}
\begin{minipage}{0.35\textwidth}
\includegraphics[angle=0,scale=0.7]{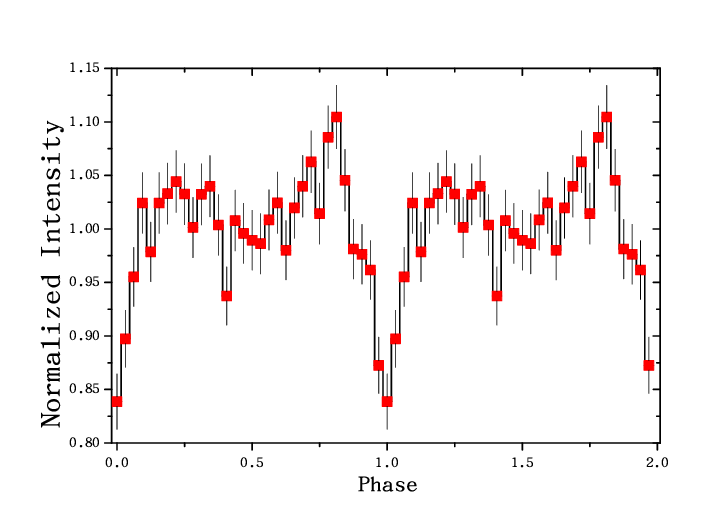}
\end{minipage}

\caption{\textit{Left}: Folded pulse profile of SMC X-2 in (3-79) keV energy range corresponding to NuSTAR observation of 2022. \textit{Right}: Folded pulse profile of SMC X-2 in (0-10) keV energy range corresponding to NICER observation of 2022. The pulse profiles are normalized about the mean count rate.}
\end{figure}

\subsection{Light curves, Pulse profiles and Pulse fraction}
We are able to investigate the characteristics of the source by utilizing NuSTAR's extensive energy coverage.
The pulse profiles are folded using a predetermined time zero-point ($T_0$) in such a way that the minimal flux bin is located at phase point = 0.0. We notice an asymmetrical character in the pulse profile pattern of the source. The folded pulse profile in (3-79) keV energy range is presented in Figure 3. The morphology of the pulse profile reveals a non-sinusoidal nature, that supports the pattern of harmonics observed in the power spectrum of the source (refer Figure 1).

The relative amplitude of the pulse profile is depicted by the pulse fraction (PF). It relates to X-ray emission from the accretion column (pulsed emission) and other regions of the accretion flow or NS surface (unpulsed emission) \citep{w}. It is defined as, $PF=\;\frac{P_{max}-P_{min}}{P_{max}+P_{min}}$ , $P_{min}$ and $P_{max}$  being the minimum maximum and intensities of pulse profile respectively. It is observed that the PF seems to increase from 5-10 keV, and is consistent with being constant within the uncertainties at higher energies. However, the pulse fraction has a diminishing tendency around specific energies, which is suggestive of absorption component identified in the X-ray spectrum of the source. For some X-ray pulsars, it has been noted in the past that the pulse fraction falls off around the CRSF energy \citep{TSY, 39, 37, y, lut}. The variation of Pulse Fraction with energy has been presented in Figure 5.

\begin{figure}

\begin{center}
\includegraphics[angle=0,scale=0.3]{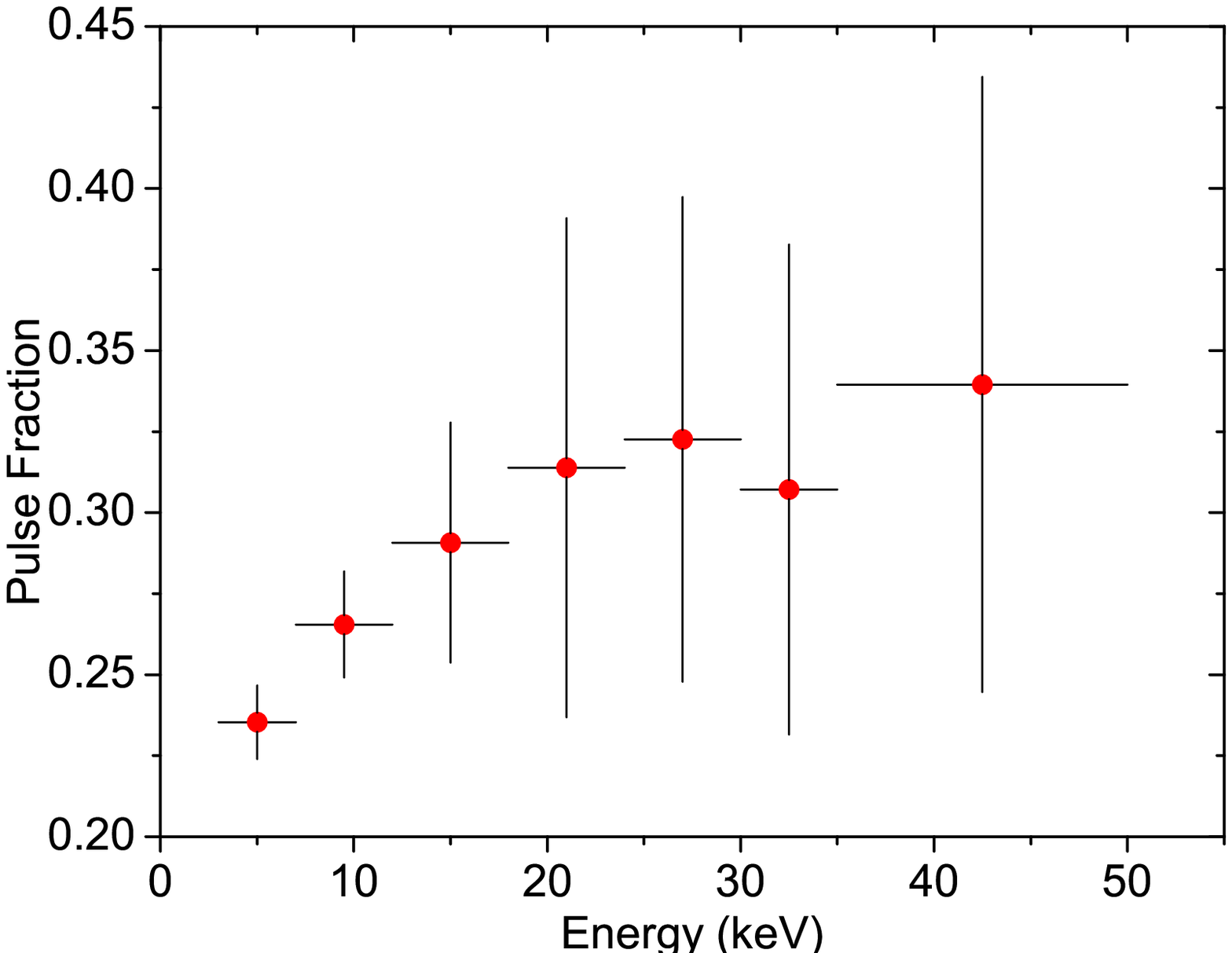}
\end{center}
\caption{Variation of pulse fraction of the source with energy using NuSTAR observation.}
\end{figure}

\section{Spectral Analysis}
The broadband joint NICER-NuSTAR spectra of SMC X-2 has been fitted using the NPEX model \citep{Mihara}. The model is termed as NPEX as it combines negative and positive power laws with a single exponential cutoff factor. The continuum spectra of X-ray pulsars may be described suitably using a continuum formula of the form :

F(E) $\propto (E^{-\alpha} + E^{\beta})\;\times \exp(-E/kT)$ ,

where $\alpha$ , $\beta$ and $T$ are positive parameters and k denotes the Boltzmann constant. The NPEX continuum has a distinct physical significance; it corresponds to a photon number spectrum for unsaturated thermal Comptonization in plasma of temperature T \citep{Sunyaev, Meszaros}. It transforms into the typical power law with a negative slope at low energies. The positive power-law term gradually takes over as the energy level rises. 

A minima of 20 counts per spectral bin were obtained by grouping the NuSTAR (FPMA and FPMB) data using the tool GRPPHA. To account for the non-simultaneity of the observations and instrumental uncertainties, NICER and NuSTAR spectra were simultaneously fitted using a CONSTANT model. In order to maintain comparable average count rates in the two instruments, we have taken care of the relative normalization factors between the two NuSTAR modules by freezing the constant factor corresponding to instrument FPMA as unity without imposing any constraint on FPMB. The constant factor for instrument FPMB was determined to be 1.011$\pm\;0.003$, which is pertinent based on the statistical data obtained. This displays a modest 1 percent uncertainty, which is consistent with \cite{Madsen}. The continuum emission of HMXB pulsars is inferred to originate due to the Comptonization of soft X-rays in the plasma above the neutron star's surface.  For the best fit spectral results, we explored a number of continuum models. We used the TBABS component with abundance from \cite{=} to estimate the absorption column density. Vern \citep{verner} was chosen as the cross-section for the TABS component. Initially, the continuum model CONSTANT*TBABS*NPEX was used to fit the X-ray spectra. The fit statistics ($\chi^{2}$ per degrees of freedom) after incorporating the aforementioned continuum model was found to be 2343.15 (1876). A GAUSSIAN component was incorporated to fit the positive residual observed in the X-ray spectra, that revealed the existence of an iron emission line at $\sim 6.36 $ keV with an equivalent width of $\sim 0.36$ keV.  The fit statistics  at this stage was found to be 2251.45 (1876). The X-ray spectrum was found to reveal a prominent absorption feature which has been appropriately fitted using the GAUSSIAN absorption component (GABS). This improved the fit statistics significantly to 1954.22 (1870) i.e., $\sim 1.04 $. The absorption feature observed at $\sim31$ keV may be interpreted as a cyclotron line. The strength and width of the absorption feature were found to be $\sim 20.32$ keV and $\sim 9.47$ keV respectively. The negative and positive power-law index are estimated to be 0.4 and 3.1 keV respectively. The absorption column density (nH) was estimated to be $\sim0.2\times 10^{22}$ cm$^{-2}$. The absorbed flux in the 1-70 keV energy range is $\sim 5.49\;\times\;10^{-10}$ erg cm$^{-2}$ s$^{-1}$ which corresponds to an X-ray luminosity of $\sim 2.24\;\times\; 10^{38}$ erg s$^{-1}$ considering the source to be at a distance of 61 kpc \citep{Hilditch}. The corresponding fit parameters have been presented in Table 2 and the spectrum is presented in Figure 6.

We have also employed the bulk and thermal Comptonization model (also known as the Becker \& Wolff (BW) model or ``bwcycl'' in the latest XSPEC version) \citep{BeckerWolff} to analyse the relevance of the observed spectral features. The application details of this model are reported in \cite{39}. The ``bwcycl'' model utilizes some specific function implementation, which is used in the XSPEC distribution. This model fitted the spectra reasonably well, demonstrating that the features are model-independent. The $\chi^{2}$ of the overall fitting is found to be 1964.70 for 1870 degrees of freedom. This model is based on the thermal and bulk comptonization of the seed photons generated owing to bremsstrahlung, cyclotron, and black-body processes occurring in the accreting plasma, and is used to explain the observed spectrum of X-ray pulsars. The broadband X-ray spectra of X-ray pulsars like 4U 0115+63, 4U 1626-67, Her X-1, EXO 2030+375, LMC X-4, and 4U 1907+09 have been effectively studied using this approach \citep{39, dai, wolff, epili, ambrosi, Tobrej}. Using this model, Becker \& Wolff (2005a,b) explained the spectrum of various sources across narrow energy ranges as well. A power-law continuum with an exponential cut-off at high energies can be used to describe the spectrum produced by the Comptonization of the seed photons. Several free parameters are incorporated in this model, including the column radius ($r_{0}$), the temperature of the electrons in the optically thin region above the accretion mound ($T_e$), the magnetic field strength (B), and the accretion rate ($\dot{M}$). Other free parameters of the model are photon diffusion parameter ($\xi$), and the Comptonization parameter ($\delta$) that represents the ratio of bulk and thermal Comptonization. The column radius ($r_{0}$) represents the radius of the hot spot on the stellar surface. The overall shape of the reprocessed radiation spectrum resulting from the bulk and thermal Comptonization of the seed photons is determined by the parameter $\delta$. These  parameters must be varied in order to fit the spectral data for a specific source for given values of the stellar mass and radius. The BW model, the Gaussian emission component, and the GABS component  were used to fit the broadband spectrum of the source. The mass and radius of the source were fixed to their canonical values of 1.4 \(M_\odot\) and 10 km respectively. The blackbody, cyclotron and bremsstrahlung normalizations were also fixed according to the standard procedure to employ the model for spectral analysis $({https://www.isdc.unige.ch/ferrigno/images/Documents/BW distribution/BW cookbook.html})$. As the contribution of the black-body source term is negligible, the normalization of the blackbody seed photon component is set to zero. Also, the normalization of the cyclotron emission seed photon component and Bremsstrahlung emission seed photon component are fixed to one. The parameters related to the cyclotron line are found to be in agreement with the measurements using the NPEX model. The line energy is estimated to be $\sim$30.89 keV using this model which is fairly consistent with that of the NPEX model (Refer Table 2). The magnetic field estimated using the BW model is (2.62$\pm$0.03)$\times$10$^{12}$ G, which is comparable with (3.46$\pm$0.06)$\times$10$^{12}$ G derived using the cyclotron line energy. The ratio of bulk to thermal Comptonization is found to be $\sim 3.6 $ while the diffusion parameter is estimated to be $\sim 0.97$. Also, the electron temperature is found to be $\sim 3.13$ keV along with an estimated column radius of $\sim110$ m. 
 
\begin{figure}
\begin{center}
\includegraphics[angle=-90,scale=0.4]{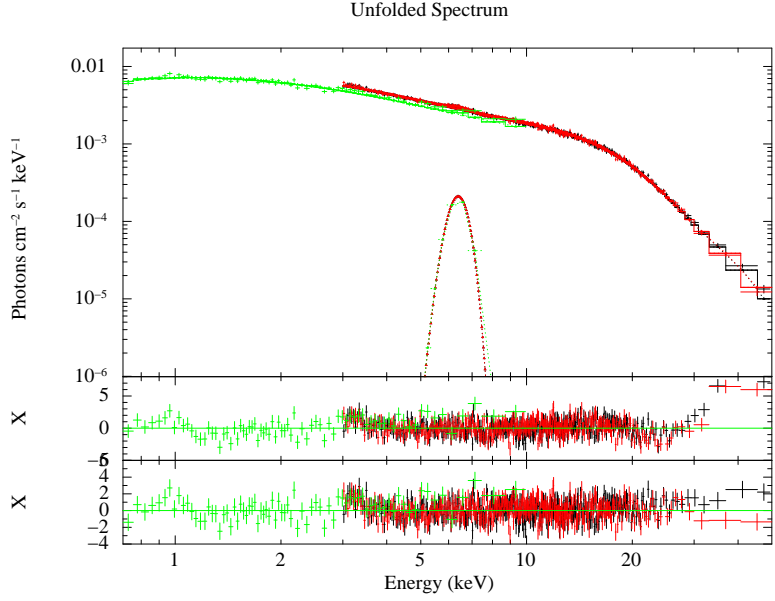}
\end{center}
\caption{The spectra corresponding to joint NICER-NuSTAR observation using NPEX model in 0.3-79.0 keV energy range. The top panel represents the unfolded spectra and the second panel represents the residuals without including GABS component. The bottom panel represents the residuals after incorporation of GABS component for fitting the residuals. The y-axis of the residual panels (middle and bottom panel) denote $\Delta\chi^{2}$ values. Black \& red colours denote NuSTAR FPMA \& FPMB spectra respectively while the green colour denotes the NICER spectra. NICER spectra is considered in 0.3-10 keV range .We neglected the spectra below 0.3 keV to counter the presence of low-energy noises while the spectra above 10 keV is skipped due to background contamination. The spectra has been rebinned for representative purpose.}
\end{figure}

\begin{table*}
\begin{center}
\begin{tabular}{ccc}
\hline										
Parameters		&		MODEL I (NPEX) 	&		MODEL II (BW)				\\
\hline														
$C_{FPMA}$		&		1(fixed)	&		1(fixed)			\\
$C_{FPMB}$		&		1.011$\pm$0.003	&		0.997$\pm$0.003	\\
$C_{NICER}$		&       0.972$\pm$0.007 &         1.023$\pm$0.003                      \\
$n_{H}\;(cm^{-2})$		&		0.20$\pm$0.01	&		0.38$\pm$0.03			\\
$\xi$ 		&		-	&		0.97$\pm$0.06					\\
$\delta$		&		-	&		3.63$\pm$0.05			-		\\
$B$ ($\times 10^ {12} G$)		&		-		&		2.62$\pm$0.03						\\
$\dot{M}$ ($10^{17}gs^{-1}$)		&		-		&		12.11$\pm$0.06						\\
$T_{e}$ (keV)		&		-		&		3.13$\pm$0.17						\\
$r_{0}$ (m)		&		-	&		110.29$\pm$5.73					\\
$\alpha$ 		&		0.42$\pm$0.05	&		-					\\
$\beta$         &        3.14$\pm$0.13   & -\\   
kT (keV)		&		3.73$\pm$0.07	&		 -					\\
$E_{Fe}$ (keV)		&		6.36$\pm$0.04	&		6.38$\pm$0.04				\\
$\sigma_{Fe}$ (keV)		&		0.36$\pm$0.05	&		0.33$\pm$0.03				\\
$E_{gabs}$		&		30.97$\pm$0.77	&		30.89$\pm$0.63					\\
$\sigma_{gabs}$		&		9.47$\pm$1.05	&		10.09$\pm$1.03				\\
$D_{gabs}$		&		20.32$\pm$5.71	&		22.15$\pm$4.97					\\
																
Flux ($\times$ $10^{-10}$ erg cm$^{2}$s$^{-1}$)		&		5.49$\pm$0.04	&		5.54$\pm$0.07			\\
$\chi^{2}_{\nu}$		&		1954.22 (1870)	&		1964.70 (1870)			\\

 \hline
 \end{tabular}
 \caption{The fit parameters of SMC X-2 using two continuum models represented by MODEL I and  MODEL II respectively. Flux was calculated within energy range (1-70) keV for the NuSTAR observation. The absorption column density (nH) is expressed in units of $10^{22} cm^{-2}$. The fit statistics $\chi_{\nu}^{2}$  denotes reduced $\chi^{2}$ ($\chi^{2}$ per degrees of freedom). $\alpha$ and $\beta$ denote the negative and positive power law indices corresponding to the NPEX model. D represents the strength of absorption line.  Errors quoted are within 1$\sigma$ confidence interval. The parameters corresponding to the BW model are denoted by $\xi$ (diffusion parameter), $\delta$ (ratio of bulk to thermal Comptonization), B (magnetic field), $T_{e}$ (electron temperature), $r_{0}$ (column radius), and $\dot{M}$ (mass accretion rate).} 
  \end{center}
 \end{table*}
 
\subsection{Phase-Resolved Spectral Analysis}
The phase-resolved spectral analysis is carried out for analyzing the anisotropic properties of the X-ray emitted by the pulsar around its rotational phase. Therefore, it would be interesting to investigate how the spectral parameters rely on the viewing angle of the NS. We have resolved the estimated pulse period of the source into 10 equal segments and generated the spectra corresponding to each segment.
\begin{table*}
\begin{center}
\begin{tabular}{ccccccc}
\hline
Phase Interval	&	$\beta$	&	$\alpha$	&	kT  (keV)	&	Fe line (keV)	&	Fe line  width (keV)	&	Flux	\\
\hline
0-0.1	&	0.26$\pm$0.07	&	2.77$\pm$0.29	&	3.57$\pm$0.19	&	6.29$\pm$0.11	&	0.35$\pm$0.14	&	4.59$\pm$0.02	\\
0.1-0.2	&	0.30$\pm$0.08	&	2.96$\pm$0.37	&	3.48$\pm$0.23	&	6.34$\pm$0.14	&	0.29$\pm$0.15	&	4.97$\pm$0.01	\\
0.2-0.3	&	0.46$\pm$0.15	&	2.03$\pm$0.44	&	4.01$\pm$0.35	&	6.34$\pm$0.15	&	0.31$\pm$0.16	&	6.10$\pm$0.03	\\
0.3-0.4	&	0.60$\pm$0.12	&	2.80$\pm$0.39	&	3.48$\pm$0.25	&	6.30$\pm$0.08	&	0.29$\pm$0.11	&	6.56$\pm$0.03	\\
0.4-0.5	&	0.44$\pm$0.13	&	2.19$\pm$0.32	&	3.93$\pm$0.25	&	6.44$\pm$0.12	&	0.40$\pm$0.14	&	5.73$\pm$0.02	\\
0.5-0.6	&	0.27$\pm$0.08	&	2.32$\pm$0.33	&	3.68$\pm$0.22	&	6.41$\pm$0.10	&	0.42$\pm$0.13	&	4.44$\pm$0.01	\\
0.6-0.7	&	0.40$\pm$0.11	&	2.26$\pm$0.33	&	3.84$\pm$0.25	&	6.42$\pm$0.11	&	0.41$\pm$0.14	&	5.85$\pm$0.03	\\
0.7-0.8	&	0.80$\pm$0.31	&	1.73$\pm$0.38	&	4.33$\pm$0.35	&	6.38$\pm$0.10	&	0.36$\pm$0.10	&	6.12$\pm$0.03	\\
0.8-0.9	&	0.55$\pm$0.10	&	2.38$\pm$0.12	&	3.71$\pm$0.37	&	6.31$\pm$0.07	&	0.34$\pm$0.10	&	5.92$\pm$0.02	\\
0.9-1.0	&	0.55$\pm$0.11	&	2.18$\pm$0.20	&	4.37$\pm$0.17	&	6.29$\pm$0.07	&	0.34$\pm$0.09	&	4.91$\pm$0.02	\\

\hline
\end{tabular}
\caption{The fit parameters of SMC X-2 along the phase intervals obtained by phase-resolved spectroscopy. Flux is estimated within energy range (3-50) keV for the NuSTAR observation and is expressed in units of $10^{-10}$ ergs cm$^{-2}$ s$^{-1}$. $\alpha$ and $\beta$ denote the negative and positive power law indices corresponding to the NPEX model.} 
\end{center}
\end{table*}
\begin{figure}

\includegraphics[angle=0,scale=0.4]{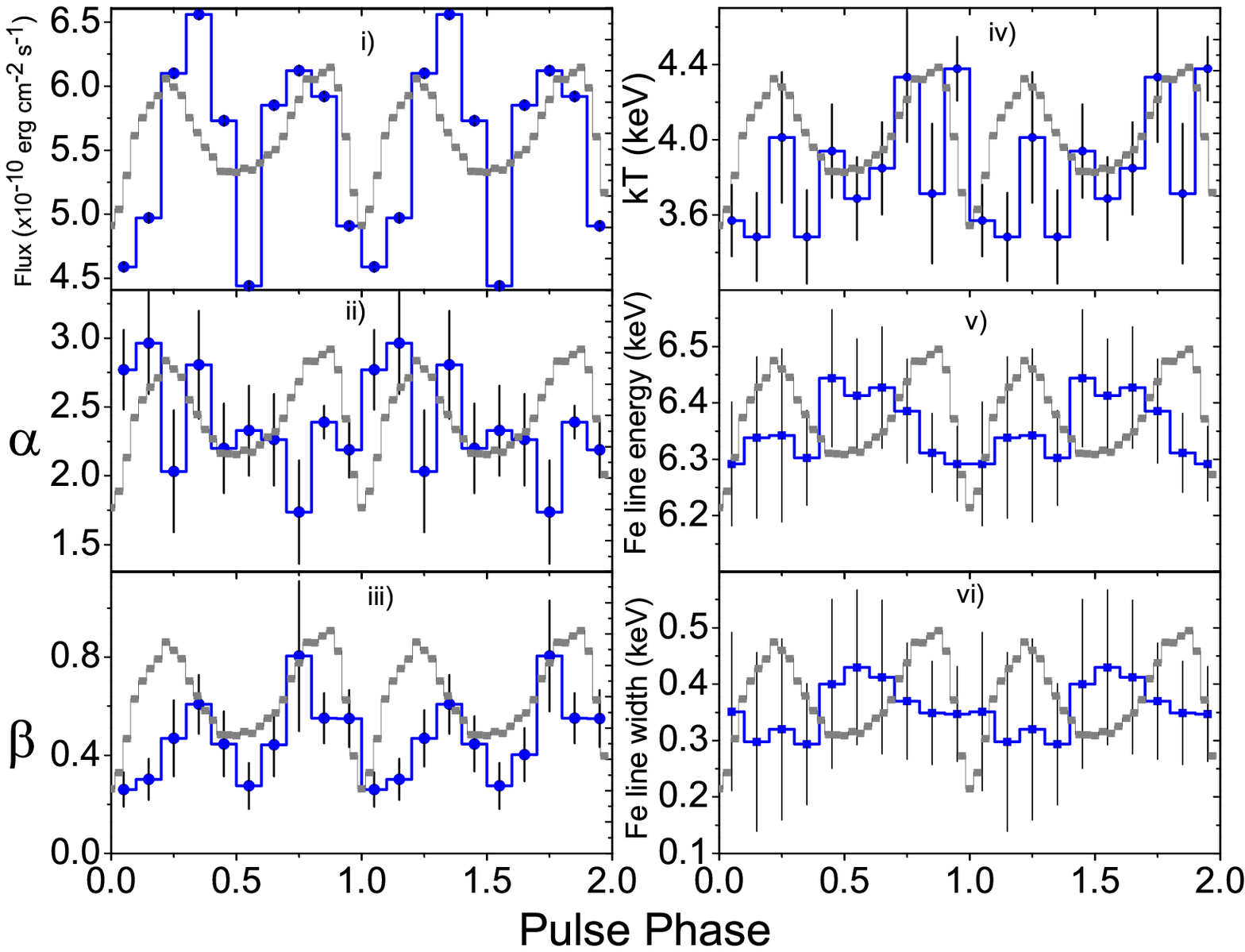}

\caption{Variation of spectral parameters:Flux (i), negative and positive power law indices ($\alpha$ (ii) and $\beta$ (iii)) plasma temperature (kT) (iv),  \& line energy (v) and width of the Fe-line (vi) with pulse phase for NuSTAR observation of SMC X-2. Roman numerals denote the positions of the spectral parameters.}
\end{figure} 
The absorption feature was observed only along the end phase bins i.e., 0.85 and 0.95. We thoroughly searched for the absorption feature in the spectra of other pulse phases, and no extra absorption lines were prominently observed. The pulse phase variation of the spectral parameters are presented in Figure 7 and the residuals corresponding to phase bin 0.95 (i.e., phase interval 0.9-1.0) revealing the detection of the absorption feature is presented in Figure 8. We have also included the residuals for phase bin 0.55 (i.e., phase interval 0.5-0.6) in Figure 8 (bottom panel) for comparison, revealing no prominent absorption feature.

\begin{figure}
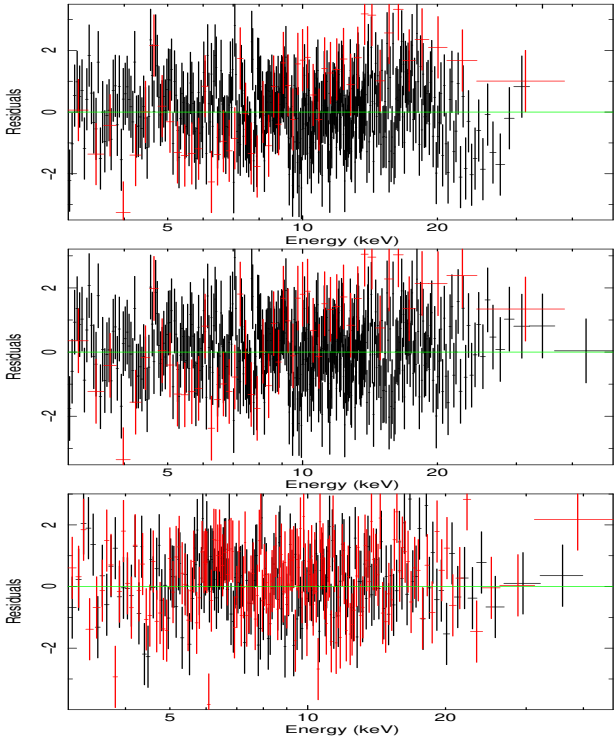

\begin{minipage}{1.75\textwidth}
\includegraphics[width=0.1\linewidth, height=0.35\textheight, angle=270]{witoutgabs}
\end{minipage}

\begin{minipage}{1.75\textwidth}
\includegraphics[width=0.1\linewidth, height=0.35\textheight, angle=270]{delchi}
\end{minipage}

\begin{minipage}{1.75\textwidth}
\includegraphics[width=0.1\linewidth, height=0.35\textheight, angle=270]{delchi6}
\end{minipage}

\caption{Top:Residuals corresponding to phase bin 0.95 where the absorption feature is observed. Middle: Residuals corresponding to phase bin 0.95 where the absorption feature is fitted using a GABS component. Bottom: The residuals for phase bin 0.55  revealing no prominent absorption feature.}
\end{figure}
The spectra corresponding to all the phase bins are well fitted using MODEL I. The spectral parameters along the phase bins is presented in Table 3. The spectral parameters are found vary  significantly relative to the pulsed phase. The absorbed flux corresponding to each phase bin is estimated in the (3-50) keV energy range. The top panel of Figure 7 represents the morphology of flux variation that is apparently similar to the nature of the pulse profile of the source. The flux ranges in the limit (4.44-6.56) $\times 10^{-10}$ ergs cm$^{-2}$ s$^{-1}$. The power law photon indices exhibit a dual-peaked profile with a phase shift relative to the pulse profile. The negative index is found to decrease near the peak emission of the profile while the positive index exhibits an increase near the peak emission indicating a softer spectrum at that phase bin. The positive index is also observed to be almost coincident adjacent to the minima of the pulse profile. The positive power law index attains a peak value of 0.8 around the phase interval 0.7-0.8. Similarly, the negative power law index attains a peak value of 2.96 and the lowest value of 1.74 along the phase intervals 0.1-0.2 and 0.7-0.8 respectively. The apparent anti-correlated behavior of $\alpha$ and $\beta$ exhibits a degeneracy as they trend slightly down and up, respectively. The two parameters degenerate and seem to compensate for one another, resulting in a steady spectral shape, which was also verified by freezing $\beta$ and allowing only $\alpha$ to vary. The plasma temperature (kT) is also found to exhibit some variation with the pulse phase and is found to range between (3.48-4.38) keV. It is found to attain maxima near the end of the primary peak in the pulse profile. The iron line parameters are also found to exhibit complex variability relative to the pulse phase. The centroid energy and width of the Fe line is found to deviate by about 7-8 percent relative to the phase-averaged value. The centroid energy of the Fe line attains a peak value of $\sim6.44$ keV near the minima or decaying phase of the pulse profile around the phase interval 0.4-0.5. The minima of the line energy ($\sim6.29$ keV) coincides with the intial and final phase bin. Similarly, the width of the Fe line is found to range in the limit (0.29-0.42)  keV. It is fairly constant over the initial and final phase intervals except in the phase interval 0.4-0.5 where an appreciable increase in the width is observed.

\subsection{Time Resolved Spectral Analysis}
\textbf{We resolved the source light curve into six segments to examine the evolution of the spectral parameters over the course of the observation.} Based upon the variability of the light curve and the  observation span, good time interval (GTI) files were  used to generate spectra for each segment. The best spectral fit was obtained using Model I. The spectral parameters along the six segments is presented in  Table 4. The source flux was found to vary in the range $\sim5.66 \;\times\;10^{-10}$ to $5.76 \;\times\;10^{-10}$ erg cm$^{-2}$ s$^{-1}$ along the six segments of the light curve. The plasma temperature (kT) is found to vary throughout the observations and exhibits a decaying trend relative to time. Similarly, the negative and positive  power law indices are found to evolve significantly with time. The negative power law index seems to exhibit an overall increasing trend while the negative power law index is found to decay relative to time \textbf{displaying a degeneracy in their behavior.} The centroid energy of the iron line and its width are  found to exhibit moderate variabilities with time. The characteristic absorption feature was distinctly observed in the X-ray spectra corresponding to all the segments. The parameters associated with cyclotron line are found to evolve slightly with time. The variations in the spectral parameters and the cyclotron line parameters have been presented in Figure 9. The variation of the cyclotron line energy relative to flux along the NuSTAR segments is presented in Figure 10.
\begin{table*} 
\begin{center}
\resizebox{\textwidth}{!}{
\begin{tabular}{ccccccccccc}

\hline
Time (MJD)	&	$\beta$	&	$\alpha$	&	KT (keV)	&	Fe line (keV)	&  Width  (keV)	&	$E_{cyc}$	&	$\sigma_{cyc}$	&	$D_{cyc}$	&	Flux & $\chi^{2}_{\nu}$	\\
\hline
59773.85	&	0.87$\pm$0.26	&	1.86$\pm$0.39	&	4.61$\pm$0.35	&	6.26$\pm$0.09	&	0.43$\pm$0.09	&	29.67$\pm$2.03	&	7.78$\pm$1.42	&	9.72$\pm$4.37	&	5.76$\pm$0.02	& 930.34(950)\\
59773.96	&	0.73$\pm$0.27	&	1.85$\pm$0.43	&	4.80$\pm$0.50	&	6.40$\pm$0.18	&	0.66$\pm$0.09	&	31.64$\pm$2.40	&	7.92$\pm$2.51	&	13.22$\pm$6.95	&	5.64$\pm$0.01	&930.62(942)\\
59774.06	&	0.39$\pm$0.14	&	2.40$\pm$0.40	&	4.47$\pm$0.48	&	6.22$\pm$0.13	&	0.48$\pm$0.17	&	31.13$\pm$2.51	&	9.07$\pm$2.95	&	20.69$\pm$8.47	&	5.66$\pm$0.02	&927.13(936)\\
59774.26	&	0.68$\pm$0.29	&	2.20$\pm$0.54	&	4.42$\pm$0.44	&	6.32$\pm$0.10	&	0.65$\pm$0.13	&	30.26$\pm$2.55	&	8.79$\pm$3.31	&	14.59$\pm$8.53	&	5.74$\pm$0.02	&891.74(910)\\
59774.47	&	0.54$\pm$0.26	&	2.22$\pm$0.22	&	4.15$\pm$0.31	&	6.28$\pm$0.10	&	0.41$\pm$0.13	&	28.96$\pm$1.54	&	5.81$\pm$1.77	&	5.06$\pm$1.42	&	5.68$\pm$0.02	&906.85(892)\\
59774.67	&	0.32$\pm$0.14	&	2.59$\pm$0.21	&	3.60$\pm$0.14	&	6.33$\pm$0.25	&	0.48$\pm$0.11	&	30.97$\pm$0.43. 	&	5.09$\pm$1.75	&	4.32$\pm$1.66	&	5.60$\pm$0.03	&986.46(945)\\

\hline
\end{tabular}}
\caption{The fit parameters of SMC X-2 along the NuSTAR segments obtained by time-resolved spectroscopy. Flux is estimated within energy range (3-50) keV for the NuSTAR observation and is expressed in units of $10^{-10}$ ergs cm$^{-2}$ s$^{-1}$. The fit statistics $\chi_{\nu}^{2}$  denotes reduced $\chi^{2}$ ($\chi^{2}$ per degrees of freedom). $\alpha$ and $\beta$ denote the negative and positive power law indices corresponding to the NPEX model. The subscript ($cyc$)  denotes the parameters corresponding to the cyclotron line.} 
\end{center}
\end{table*}

\begin{figure}
\begin{minipage}{0.15\textwidth}
\includegraphics[angle=270,scale=0.25]{lcurve_rev}
\end{minipage}
\hspace{0.2\linewidth}
\begin{minipage}{0.15\textwidth}
\includegraphics[angle=0,scale=0.25]{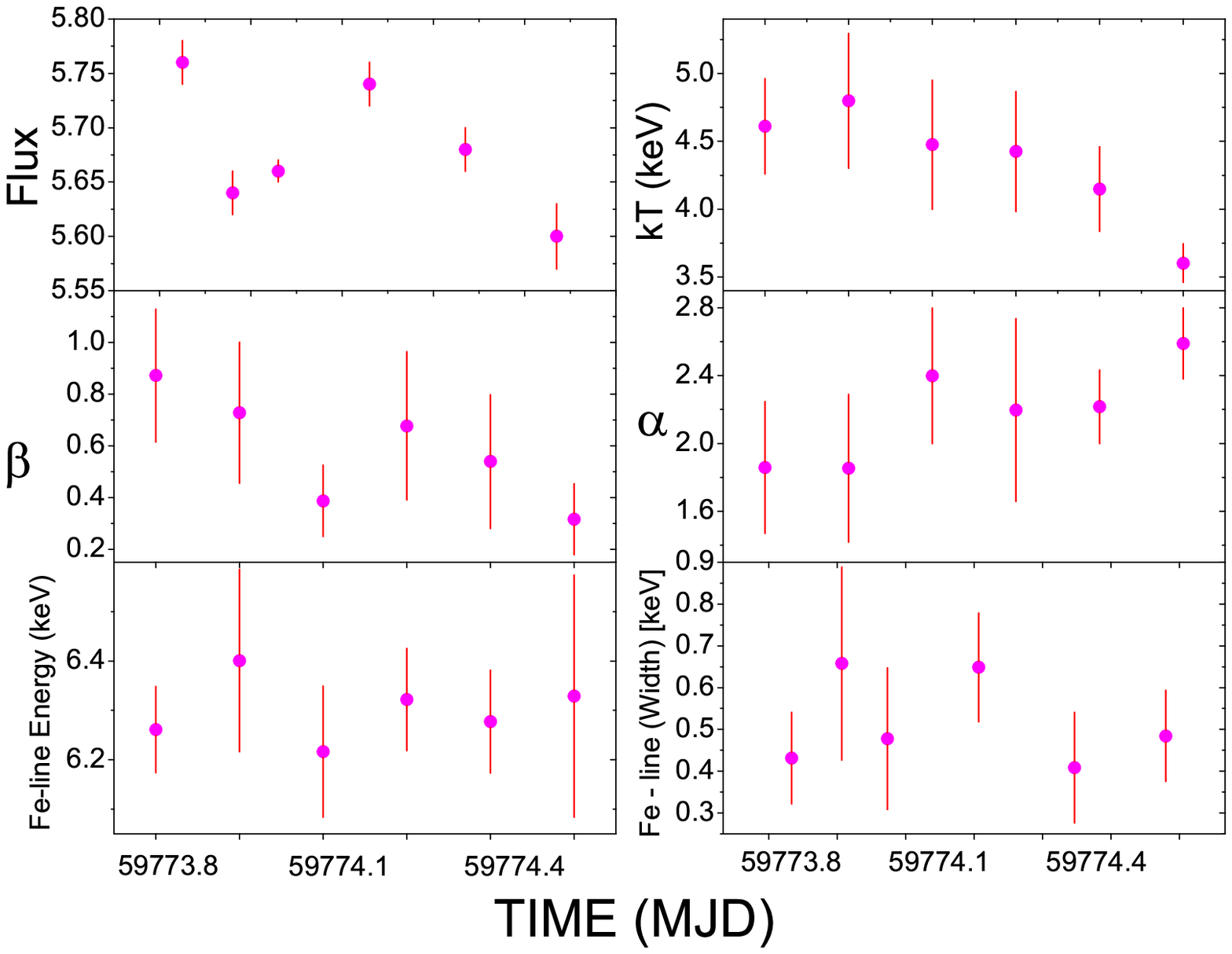}
\end{minipage}
\hspace{0.2\linewidth}
\begin{minipage}{0.15\textwidth}
\includegraphics[angle=0,scale=0.25]{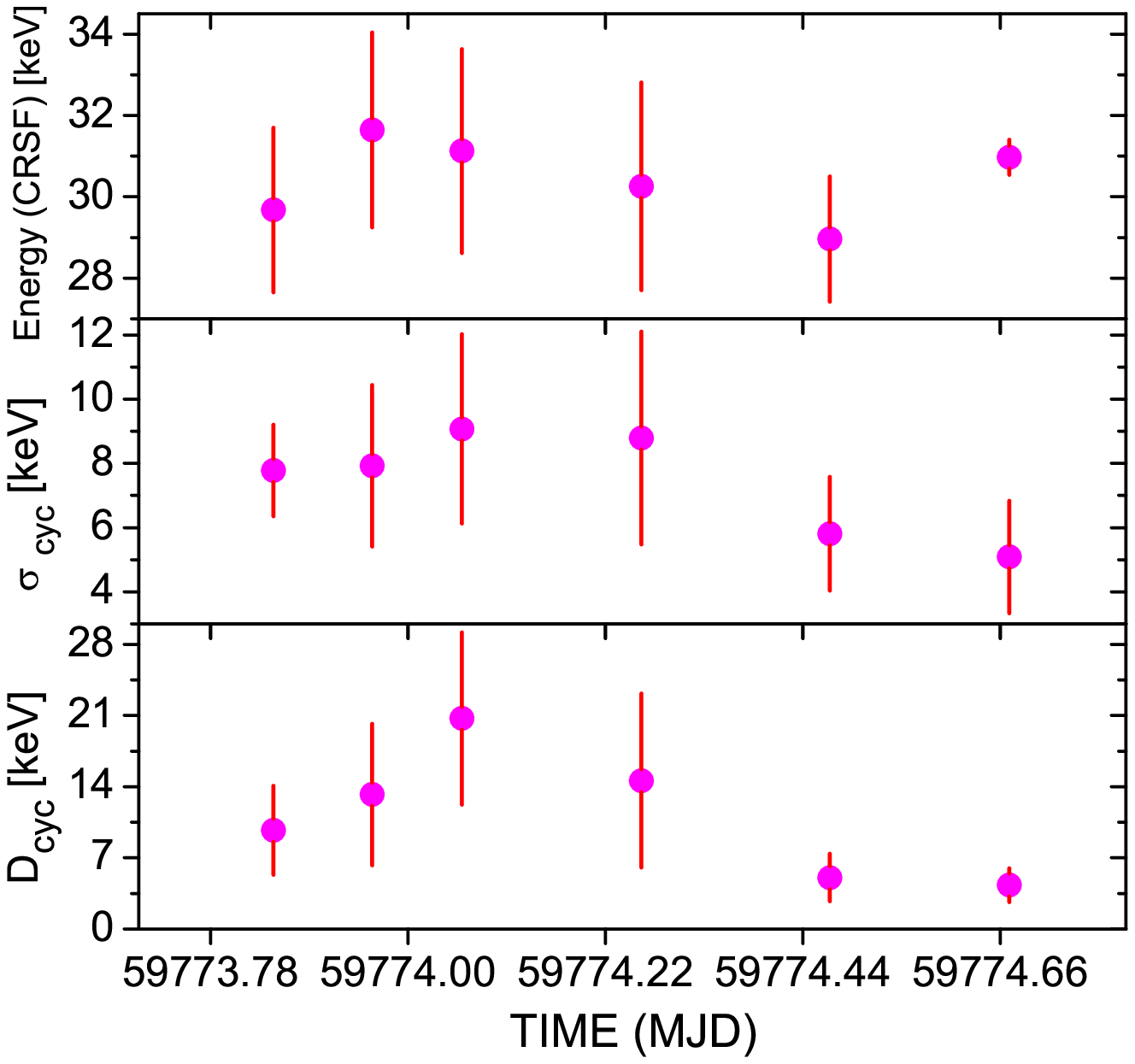}
\end{minipage}
\caption{\textit{Left}: NuSTAR light curve of the source. \textit{Middle}: Variation of spectral parameters with time along the NuSTAR segments of SMC X-2. Flux (in units of $10^{-10}$ ergs cm$^{-2}$ $s^{-1}$), negative and positive power law indices, plasma temperature ,  \& line energy and width of the Fe-line. \textit{Right}: Variation of cyclotron line parameters.}
\end{figure}
\begin{figure}

\includegraphics[angle=0,scale=0.33]{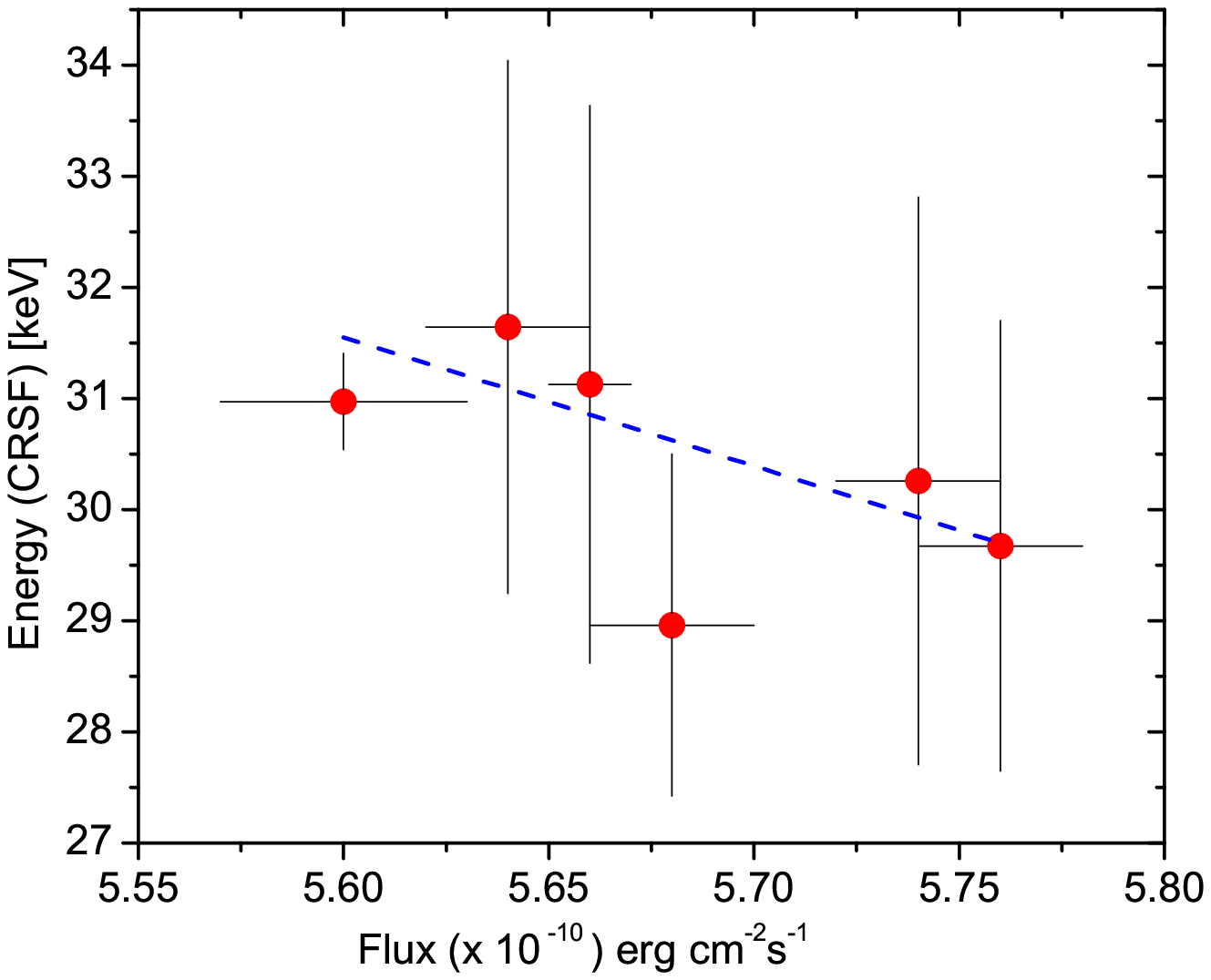}

\caption{Variation of cyclotron line energy with flux for NuSTAR observation of SMC X-2 along all the segments obtained by time-resolved spectroscopy. The dashed line represents the best fitted line characterized by Pearson's correlation coefficient of -0.64.  The variance of Pearson's coefficient is $\sim$ 41\%. Errors are within 1$\sigma$ confidence interval.}
\end{figure}

 \section{Discussion \& Conclusion}
The spectral and temporal characteristics of this HMXB source have been examined using the joint NuSTAR and NICER data observed on 13-07-2022. The coherent X-ray pulsation of the source  estimated the pulse period of the source to be  2.37281(3) s. It is observed that the pulse profiles are asymmetric, which is typical of X-ray pulsars. Several theoretical models have been used to support the asymmetrical nature of the pulse profiles. The distorted magnetic dipole field, where the magnetic poles are not exactly opposite to one another, has been cited as a potential cause for the asymmetry of pulse patterns in the reports of \cite{103,104,105,106}. The asymmetric structure of pulse profiles \citep{y,101,z} has also been identified as possibly being caused by an asymmetric accretion stream.  As the X-ray luminosity of the source is high, a fan-beam pattern is expected.  
 
The PF represents an intriguing aspect of the source and is the fraction of photons that contribute to the modulated portion of the flux. The PF apparently increases at lower energies, and is consistent with being constant within the uncertainties at higher energies. Several X-ray pulsars are known to exhibit a non-monotonic dependence of the PF on energy around the cyclotron line \citep{38,37,39}. 
  
The X-ray spectrum has been well fitted by using the NPEX model along with GAUSSIAN and GABS components. In addition to an iron emission line at $\sim 6.4$ keV, we confirm the presence of a CRSF at $\sim 31$ keV. We confirm the existence of CRSF at $\sim 31$ keV which is relatively higher than the previous measurements of line energy ($\sim$ 27 keV) during the 2015 outburst of the source \citep{Jaisawal}. CRSFs are significant characteristic absorption features that have been observed in the X-ray continuum of accreting X-ray pulsars. CRSFs are known to be formed due to the resonant scattering of hard X-ray photons and electrons in quantized energy states \citep{Meszaros}. The energy levels known as Landau levels, are equally spaced and depend on the  strength of the magnetic field. The separation between the Landau levels represents the centroid energy of the cyclotron feature. The centroid energy of the cyclotron line is directly related with the magnetic field strength as $E_{cyc}$ = 11.6 $\times\; B_{12} \times\; (1 + z)^{-1}$ (keV), where $ B_{12}$ is the magnetic field scaled in units of $10^{12}$ G and z ($\sim0.3$)represents the gravitational redshift in the scattering region for standard NS parameters. Therefore, the detection of cyclotron line is an efficient method for estimating the magnetic field of accretion powered X-ray pulsars. The magnetic field corresponding to the CRSF at $\sim 31$ keV is estimated to be $\sim 3.47 \times\;10^{12}$ G.

 The physical BW model reveals that the bulk Comptonization dominates the thermal Comptonization which may be expected in the high luminosity states. As the plasma traverses the radiative shock, a substantial compression may occur, making it ideal for the bulk Comptonization of seed photons. During this  process, particles may exhibit a gain in mean energies if
the collision with the scattering centers are involved in a converging flow \citep{Laurent, Turolla}. Except at the greatest photon energies, bulk Comptonization prevails over thermal Comptonization as the inflow velocity of the electrons in an X-ray pulsar accretion column is significantly faster than their thermal velocity \citep{Titarchuk}.

The critical luminosity ($L_{c}$) which marks the transition in the accretion states is approximated using the relation:
$L_{c} = 1.5 \times10^{37}B_{12}^{16/15}$ erg s$^{-1}$ \citep{Beckerr}. The critical luminosity corresponding to the measured magnetic field is estimated to be $\sim 5.66 \times10^{37}$ erg s$^{-1}$. It is evident that the estimated luminosity is higher than the critical luminosity. Therefore, it may be inferred that the source may be accreting in the super-critical regime. 

Phase-resolved spectroscopy has also been employed to test the existence of distinct spectral features. Various sources are known in which the cyclotron line or its higher harmonics have only ever been found during particular rotational phases. In sources like Vela X-1 \citep{+} and KS 1947+300 \citep{-}, the fundamental line is occasionally only detectable during specific pulse phases. This implies that the contributions of the two accretion columns are varied and/or that the emission pattern during the pertinent pulse is such that the CRSF is extremely shallow or filled by spawned photons \citep{*}. Significant variation in the spectral characteristics may be seen in the pulse phase resolved spectroscopy with 10 equally spaced phase bins. Only along the phase bins 0.85 and 0.95, the absorption feature was distinctly observed, and it was less noticeable in some phases. A significant gradient in the strength of the magnetic field over the visible column height or latitudes on the surface of the NS \citep{99,x} may explain the lack or non-detection of the absorption line in the pulse phases. High mass accretion rates lead to the appearance of  accretion columns that are supported by high internal radiation pressure and are constrained by the strong magnetic field \citep{Basko, Wang Frank, Mushtukov}. Thus, the cyclotron line may originate from the accretion column \citep{Nishimura, Nishimuraa, Schonherr} or it may be caused by the atmosphere of the neutron star reflecting X-rays \citep{Poutanen, Lutovinovv}. The scattering feature may disappear from the measured energy spectra if there is a significant gradient in the magnetic field intensity over the visible column height or latitudes on the stellar surface. However, it is conceivable for the viewer to only see a portion of the accretion column at some phases of the pulse because it is partially eclipsed by the neutron star \citep{Mushtukovv}. As observed here, the cyclotron line can appear during some phases of pulsations because of the comparatively limited dispersion of the magnetic field intensity over the visible portion of the column. It was observed that the power-law indices exhibit notable variabilities with the pulse phase of the system. 

We have used time-resolved spectroscopy to investigate the  spectrum evolution of the pulsar. 
The plasma temperature (kT) exhibits a downward trend while $\alpha$ trends slightly down and $\beta$ trends slightly up,  displaying a degeneracy in their anti-correlated behavior. The two parameters degenerate and seem to compensate for one another,
resulting in a steady spectral shape. We validated the same by freezing $\beta$, allowing only $\alpha$ to vary and observed that the continuum exhibits a steady spectral shape relative to time. It is also evident that the width of th absorption line is  constant within the uncertainties, while the centroid energy and depth exhibit a significant deviation only along the last time bin. It is observed that the cyclotron line energy is moderately anti-correlated with the X-ray flux suggesting that the source may be accreting in the super critical regime. In this regime, the accreting material slows down near the surface of the neutron star, creating a radiation-dominated shock in the accretion column where radiation pressure is high enough to stop the accreting material above the NS. In such a case, the sidewalls of the accretion column release X-rays in the form of a fan-beam \citep{Davidson}. As the accretion rate increases, the shock area ascends higher in the accretion column, where the magnetic field is weaker. \cite{Burnard} formulated a straightforward theory regarding the underlying physical cause of the anti-correlation. According to \cite{Basko}, the height of the radiative shock above the NS
 surface and, consequently, the region where lines are formed should increase linearly with increasing accretion rate. They pointed out that this would indicate a decrease in field strength and, consequently, a decrease in the cyclotron line energy. Various pulsars like 4U 0115+63 \citep{Nakajima} and V 0332+53 \citep{y} have been known to exhibit such an anti-correlation. The strength of the  magnetic field and cyclotron line energy should vary according to the following law with variation in  mass accretion rate:

 $E_{cyc}$($\dot{M}$) = E $\times\left(\frac{R}{H(\dot{M}) + R}\right)^{3}$,

where E represents the line emitted from the NS surface, R is the radius of the NS, and H denotes the actual height of the line forming region above the neutron star surface.


\section*{Acknowledgements}
This work was carried out utilising the NuSTAR and NICER data provided by NASA High Energy Astrophysics Science Archive Research Center (HEASARC), Goddard Space Flight Center. MT acknowledges CSIR for the research grant 161-411-3508/2K23/1. We appreciate the research facilities provided by the IUCAA Centre for Astronomy Research and Development (ICARD), Department of Physics, NBU. We would like to thank the honourable reviewer for the insightful suggestions. 

\section*{Data availability}
 
The HEASARC data archive provides access to the observational data utilised in this study, which is open to the public for research.

\bibliographystyle{elsarticle-harv}

\end{document}